\documentclass[3p, review, 11pt]{elsarticle}
\usepackage{flushend}
\usepackage{epsfig}
\usepackage{amsmath}
\usepackage{flushend}
\usepackage{multirow}
\usepackage{color}
\usepackage{enumerate}
\usepackage{easylist}

\usepackage{amsthm}
\usepackage{amssymb }
\usepackage{amsmath}
\usepackage{bm}
\usepackage{float}
\usepackage{mathrsfs}
\usepackage{amsfonts}
\usepackage{graphicx}
\usepackage{algorithm, algpseudocode}
\usepackage{subfigure}
\def\u{u}

\def\n{(n)}
\def\msr{m}  % measurement
\def\msrV{\bm{m}} % measurement vector
\def\vm{m_v} % Voluntary motion %\def\vm{m_v}
\def\im{m_i} % Involuntary (Tremor) motion
\def\wl{L} % Window length
\def\ld{J} % Level of Decomposition
\def\po{p} % prediction order
\def\gt{^{\text{(GT)}}} % Ground Truth
\def\j{_{(j)}} % Index
\def\a{\bm{a}}
\def\p{\bm{p}}
\def\T{\bm{T}}

\def\w{\bm{w}}
\def\fd{f_d}

\def\s{\bm{s}}
\def\volpr{\hat{\bm{m}}} % Voluntary movement prediction

\def\P{\bm{P}}
\def\S{\bm{S}}

\def\WAKE{\text{WAKE}}
\def\HPA{\text{HPA}}
\def\RTP{\text{RTP}}
\def\S{\bm{S}}

\def\z{\bm{z}}
\def\br{\bm{r}}
\def\hx{\hat{\bm{x}}}

\def\K{\bm{K}}
\def\Q{\bm{Q}}
\def\P{\bm{P}}

\def\x{\bm{x}}
\def\A{\bm{A}}

\def\F{\bm{F}}

\def\X{\bm{X}}

\def\k{(k)}
\def\pk{(k-1)}
\def\kk{(k|k)}
\def\kpk{(k|k-1)}
\def\pkpk{(k-1|k-1)}
\def\z{\bm{z}}

\algnewcommand\Input{\item[\hspace{6pt}\textbf{Input:}]}
\algnewcommand\Output{\item[\hspace{6pt}\textbf{Output:}]}
\algnewcommand\OutputVal{\textbf{output} }
\begin{document}

\begin{frontmatter}

\title{$\WAKE$: Wavelet Decomposition Coupled with Adaptive Kalman Filtering for Pathological Tremor Extraction}

\author{Soroosh Shahtalebi$\dag$, Seyed Farokh Atashzar$\ddag$, and Rajni V. Patel$\ddag$, Arash Mohammadi$\dag$}

\address{$\dag$Concordia Institute for Information Systems Engineering,  Concordia University,\\
1455 De Maisonneuve Blv. W., EV-009.187, Montreal, QC, Canada, H3G-1M8\\
$\ddag$ Electrical and Computer Engineering, University of Western Ontario, London, ON, Canada\\
Emails:' $\{$s\_shahta, arashmoh$\}$@encs.concordia.ca, $\{$satashza, rvpatel$\}$@uwo.ca}

\begin{abstract}
%=============================================
Pathological Hand Tremor (PHT) is among common symptoms of several neurological movement disorders, which can significantly degrade quality of life of affected individuals. Beside pharmaceutical and surgical therapies, mechatronic technologies have been utilized to control PHTs. Most of these technologies function based on estimation, extraction, and characterization of tremor movement signals. Real-time extraction of tremor signal is of paramount importance because of its application in assistive and rehabilitative devices. In this paper, we propose a novel on-line adaptive method which can adjust the hyper-parameters of the filter to the variable characteristics of the tremor. {\color{black} The proposed \textit{Wavelet decomposition coupled with Adaptive Kalman filtering technique for pathological tremor Extraction, referred to as the WAKE  framework}}, is composed of a new adaptive Kalman filter and a wavelet transform core to provide indirect prediction of the tremor, one sample ahead of time, to be used for its suppression. In this paper, the design, implementation and evaluation of WAKE are given. {\color{black} The performance is evaluated based on three different datasets, the first one is a synthetic dataset, developed in this work, that simulates hand tremor under ten different conditions. The second and third ones are real datasets recorded from patients with PHTs.} The results obtained from the proposed WAKE framework demonstrate significant improvements in the estimation accuracy in comparison with two well regarded techniques in the literature.
%=============================================
\end{abstract}


\end{frontmatter}
%OOOOOOOOOOOOOOOOOOOOOOOOOOOOOOOOOOOOOOOOOOOOOOOOOOOOOOOOO
\section{Introduction} \label{sec:Introduction}
%OOOOOOOOOOOOOOOOOOOOOOOOOOOOOOOOOOOOOOOOOOOOOOOOOOOOOOOOO
{\color{black}The number of seniors over the age of 60 is expected to increase from 841 million in 2013 to more than 2 billion by 2050~\cite{UN2013}}. This directly increases the occurrence of age-related neurological disorders such as Parkinson's Disease (PD), Essential Tremor (ET), and their common motor symptoms such as Pathological Hand Tremor (PHT)~\cite{Kotsavasiloglou:2017}\nocite{Camara:2015}-\cite{Smits:2017}. Tremor is defined as a non-volitional and pseudo-rhythmic movement which affects coordination, targeting, and speed of movements~\cite{dick2016features}\nocite{tarsy2008handbook}-\cite{Lui:2016}.  Pathological tremor can severely affect the quality of life of individuals and reduce their ability to perform the activities of daily living (ADLs). As a result, on-line and accurate estimation/extraction of PHT is of paramount importance to develop new assistive and rehabilitation technologies such as robotic rehabilitation systems and assistive robotic surgery. The extracted signals can be used in the design of technologies that aim to damp, counteract or compensate for the tremor. This may be done to assist or to rehabilitate motor control of patients using interactive robotic or wearable exoskeleton platforms.

There has been a recent surge of interest~\cite{Hossen:2010}\nocite{Atashzar:2016,Xiao:2016,Taheri:2015,Veluvolu:2013, giuffrida2009clinically,Kucukelbir:2009, Dosen:2015,kiguchi2012emg, Bo:2011, veluvolu2011estimation, Wang:2014}-\cite{Veluvolu:2010} in development of efficient algorithms that analyze motion signals (that are composed of involuntary components, i.e., tremors, and voluntary components) to accurately extract pathological tremors. In this regard, various modalities of motion have been studied, including angular and linear velocities and accelerations. In addition, various sensors have been utilized to measure the motions such as gyroscopes, linear accelerometers, and optical motion sensors~\cite{giuffrida2009clinically}.
Examples of the above-mentioned tremor estimation algorithms can be found in~\cite{Atashzar:2016}\nocite{Xiao:2016,Taheri:2015,Veluvolu:2013,giuffrida2009clinically}-\cite{Kucukelbir:2009}, and references therein. Among these examples, techniques that are designed based on the formulation of Fourier Linear Combiner (FLC)~\cite{Wang:2014}\nocite{Veluvolu:2010,Veluvolu:2010v1,wang2014adaptive,veluvolu2007bandlimited}-\cite{veluvolu2010estimation} are commonly used in the design of tremor-compensation mechatronic technologies. This may be attributed to the fact that FLC-based algorithms introduce low latency to the system and can be very accurate for rhythmic signals. Several variations of FLC-based methods have been proposed in literature among which the Band-limited Multiple FLC (BMFLC)~\cite{veluvolu2007bandlimited,veluvolu2010estimation} is widely considered as one of the most successful extensions. The BMFLC tracks the frequency content of the involuntary motion (i.e., tremor) within a predefined frequency band of tremor by adaptively computing the weights for each frequency. Extraction is then performed by reconstruction of the tracked motion in the time-domain. Recently in ~\cite{Atashzar:2016}, Atashzar~\textit{et al.} proposed an enhanced version of the conventional BMFLC framework (referred to as E-BMFLC) where the technique: (i) Models the complete motion while considering full range of frequency (in contrast to the conventional implementation of the BMFLC that only models the tremor frequency band), and; (ii) Modulates the memory of the embedded recursive filter. The aforementioned modifications introduced by the E-BMFLC, have resulted in statistically-significant improvement of both accuracy and consistency  in extracting PHTs for patients with PD and ET.

Despite successful implementation of the FLC-based techniques (in particular E-BMFLC), the non-stationary and variable nature of \textit{pathological} tremor, challenges accurate and real-time extraction of the signal specifically when knowledge about the expected frequency context is unavailable or uncertain. This motivates development of new and innovative signal processing solutions which can significantly enhance the performance of rehabilitative and assistive tremor-compensation mechatronic technologies, in practice. We believe the key to achieve this goal is to go beyond the paradigm of the FLC framework, which is limited to the frequency domain. The suggested solution in this paper is to conduct tremor extraction through joint incorporation of temporal and spectral features.
Although some recent research studies~\cite{hossen2010discrimination}\nocite{geman2012using,popovic2008extraction,ai2011classification,sushkova2015time}-\cite{elble2017assessment}
have considered  spectra-temporal features to analyze the body movement signals, to best of our knowledge, all existing examples are off-line techniques that only use the signal in a batch mode. The paper addresses this gap by developing an on-line wavelet-based framework to estimate the tremor signal.

In this paper, by capitalizing on the aforementioned need for development of new on-line and real-time tremor estimation/extraction solutions which incorporate spectra-temporal movement information, we propose a novel framework referred to as Wavelet Adaptive Kalman Tremor Extraction ($\WAKE$), which decomposes the signal into several spectra-temporal components via wavelet transforms and then provides myopic predictions of the tremor by means of an adaptive auto-adjustable Kalman filtering (KF) framework.
The $\WAKE$ framework incorporates a multi-rate scheme to extract the tremor out of the measurement signal by utilizing KF and wavelet transforms in an adaptive, iterative, and optimized fashion. The proposed $\WAKE$ framework consists of the following two schemes running in parallel but with different rates: (a) \textit{Real-time Tremor Prediction ($\RTP$)}, which operates in an on-line manner and provides predictions for the tremor signal in the next time instance, and; (b) \textit{Hyper Parameter Adjustment ($\HPA$)}, which operates in slower rate and optimizes hyper-parameters of the filter to boost the performance and accuracy of the $\RTP$ scheme.

In summary, the proposed $\WAKE$ framework runs in real-time and provides myopic (one-step ahead) predictions of the pathological tremor signal via an adaptive and multi-rate mechanism making it suitable for implementation on hardware and real world applications.
The $\WAKE$ framework provides several benefits including: (i) The ability to analyze the measurement signal in real-time; (ii) Providing myopic predictions for the tremor signal, and; (iii) Adaptively adjusting to the behavior changes of the tremor signal via implementation of an auto-adjusting scheme which runs in parallel to the prediction scheme.

The rest of the paper is organized as follows: Section~\ref{sec:problem} formulates tremor estimation/extraction problem, and provides a brief description of the required background materials. In Section~\ref{sec:method}, we present the proposed $\WAKE$ framework. Section~\ref{sec:exp} provides the results of the pre-setup experiments together with that of applying the $\WAKE$ on three tremor datasets. Finally Section~\ref{sec:conc} concludes the paper.

%OOOOOOOOOOOOOOOOOOOOOOOOOOOOOOOOOOOOOOOOOOOOOOOOOOOOOOOOO
\section{Problem Formulation} \label{sec:problem}
%OOOOOOOOOOOOOOOOOOOOOOOOOOOOOOOOOOOOOOOOOOOOOOOOOOOOOOOOO
\textcolor{black}{Throughout the paper, we use the following notation:  non-bold letter $x$ denotes a scalar variable,  lowercase bold letter $\x$ represents  a vector, and capital bold letter $\X$ denotes a matrix.} In this section, we briefly provide the mathematical background required for development of the proposed $\WAKE$ framework.  We consider the following model for the hand movement of a patient with pathological tremor
\begin{equation}\label{eq:CSP-Cov}
\msr\n = \vm\n + \im\n,
\end{equation}
where $\msr\n$ represents the composite motion signal (measurement); $\vm\n$ and $\im\n$ are the voluntary and involuntary (tremor) components of the motion, respectively; and $n$ denotes the time index. In the proposed $\WAKE$ framework, we use wavelet transforms to obtain multi-scale decomposition of the measurement $\msr\n$. This is similar in nature to the Fourier transform used in the FLC-based tremor extraction methodologies. However, unlike sinusoidal functions, a wavelet and the ones generated from it are localized in space, therefore, providing a mechanism to approximate $\msr\n$ by a series of scaled and translated versions of these localized functions. In other words, the wavelet transform decomposes the signal into different scales/resolutions. Lower scales provide more details of high frequency components while higher scales provide overall features associated with low frequency components. Before presenting the proposed $\WAKE$ framework, we briefly outline the wavelet decomposition technique.

%=========================================================
\subsection{Spectra-Temporal Signal Analysis}\label{sec:wavelet}
%=========================================================
Various spectra-temporal signal analysis methods have been investigated in literature including the Short Time Fourier Transform (STFT), Wigner-Ville Transform (WVT), Choi-Williams Distribution (CWD), and Wavelet Transform (WT)~\cite{Guido:2017,Rioul:1991}, among which the latter is more favorable due to providing high resolution for high-frequency signals.

A wavelet transform represents a signal into different scales and dilations of a ``finite-length" and ``fast-decaying'' oscillating waveform known as the wavelet function (mother wavelet) and scaling function (father wavelet). Mathematically speaking, a mother wavelet $\psi(t)$ should be a square integrable function, and satisfy the admissibility condition and the regularity condition, which requires $\psi(t)$ to be fast decaying or be non-zero only on a finite interval.
To form the wavelet transform of a given signal, different dilations and scales of the mother wavelet function $\psi_{j, \u}(t)$ are applied to the signal~as
\begin{equation}
\psi_{j,\u}(t) = s_{0}^{-j/2}\psi(s_{0}^{-j}t - \u\tau_{0}),\quad j,\u \in \mathbb{Z},
\end{equation}
where $\u$ and $j$ represent  dilation, and scale of the mother wavelet function, respectively. Terms $s_{0}>1$ and $\tau_{0}$ are fixed dilation and translation factors~\cite{Rioul:1991}. The discrete wavelet transform (DWT) is then defined by
\begin{equation}
\T_{x}(j,\u) = \int_{-\infty}^{\infty} \x(t)\psi^{*}_{j,\u}(t) dt.
\end{equation}
If the family of wavelets $\psi_{j,\u}(t)$ form an orthogonal basis, the signal $\x(t)$ could be recovered from its discrete wavelet decomposition ($\T_{x}(j,\u)$) as follows
\begin{equation}
\x(t) = \frac{1}{c_{\psi}}\sum_{j,\u \in Z}\T_{x}(j,\u)\psi_{j,\u}(t).
\end{equation}
The father (scaling) function, which has also the same shape as of the mother wavelet, represents smoothed image of the signal and is defined~as follows
\begin{equation}
\phi_{j,\u}(t) = s_{0}^{-j/2}\phi(s_{0}^{-j}t - \u\tau_{0}),\quad j,\u \in \mathbb{Z},
\end{equation}
where $\int_{-\infty}^{\infty} \phi_{0,0}(t) = 1$, and $\phi_{0,0} = \phi$. The mother wavelet acts as a high-pass filter, so its coefficients ($\T_x(j,\u)$) represent the \textit{details} of the signal, while the father wavelet behaves as a low-pass filter which provides the \textit{approximations} of the signal. To derive the approximation coefficients, $\S_x(j,\u)$, the father wavelet function is  convolved with the signal as
\begin{equation}
\S_{x}(j,\u) = \int_{-\infty}^{\infty} \x(t)\phi_{j,\u}(t)dt.
\end{equation}
The approximation coefficients at a specific scale $j$ represent the discrete approximation of the signal at that scale. A continuous approximation of the signal at scale $j$ is derived by summing a sequence of father wavelets at this scale factored by the approximation coefficients as follows
\begin{equation}
\hat{\x}_{j}(t) = \sum_{\u=-\infty}^{\infty} \S_x(j,\u)\phi_{j,\u}(t),
\end{equation}
where $\hat{\x}_{j}(t)$ is a smooth, scaling-function-dependent version of the signal $\x(t)$ at scale index $j$.  This completes a brief background formulation, next, we present the proposed $\WAKE$ framework.

%OOOOOOOOOOOOOOOOOOOOOOOOOOOOOOOOOOOOOOOOOOOOOOOOOOOOOOOOO
\section{The $\WAKE$ Framework} \label{sec:method}
%OOOOOOOOOOOOOOOOOOOOOOOOOOOOOOOOOOOOOOOOOOOOOOOOOOOOOOOOO
In this section, we present the proposed framework which is designed to extract the tremor signal from the raw measurements consisting of both voluntary and involuntary components of motions.
To accurately predict the tremor in the next time instant and in real-time, we propose the $\WAKE$ framework which consists of the following two multi-rate filtering schemes operating in parallel and cooperatively provide a self-adjustable tremor extraction engine:

\vspace{.025in}
\noindent
\textit{1. Real-time Tremor Prediction (RTP) Scheme}, which processes the raw sensory data in real-time and predicts the value of the voluntary motion in the next time instant.

\noindent
\textit{2. Hyper-Parameter Adjustment (HPA) Scheme}, which operates in slower rate than the $\RTP$~scheme and performs post-processing on the previous time samples to derive optimal values for the hyper-parameters. The optimized parameters provided by the $\HPA$ scheme result in higher performance in tremor extraction and prediction.

\vspace{.025in}
\noindent
In the following subsections, we elaborate on the details of the $\HPA$ and the $\RTP$ schemes, respectively.

%=========================================================
\subsection{Hyper-Parameter Adjustment (\HPA) Scheme}
%=========================================================
%---------------------------------------------------------------------
%
\begin{figure}[t!]
\centering
\includegraphics[scale=0.63]{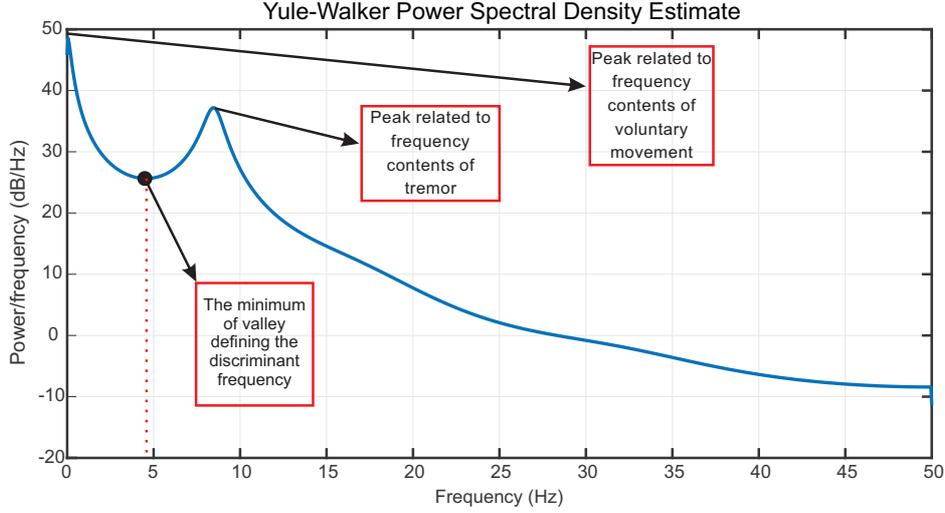}
\caption{The power spectrum estimation for a sample measurement signal containing voluntary movement and action tremor. \label{PSD}}
\end{figure}
%
%---------------------------------------------------------------------

In the proposed multi-rate $\WAKE$ framework, the $\HPA$~scheme has slower rate of execution in comparison to the $\RTP$ as it performs post-processing on the previous data samples and extracts a number of hyper-parameters which are fed into the $\RTP$~scheme. In particular, the $\HPA$~scheme performs the analysis over a pre-defined window of length~$\wl$. At each time index $n$, the last $\wl$ measured samples are combined in a vector denoted by
\begin{eqnarray}\label{eq:8}
\msrV_\wl\n &\triangleq& [\msr(n-\wl+1), \ldots,  \msr\n]^T,
\end{eqnarray}
where superscript $T$ denotes transpose operator. \textcolor{black}{Please note that Term $\msrV_\wl\n $ in Eq.~\eqref{eq:8} is a vector of ($L \times 1$) dimension, as it is constructed by stacking  the last $L$ measured samples ($\msr(n-\wl+1)$ to $\msr(n)$).} The power spectrum estimation of the windowed signal $\msrV_\wl\n$ is then computed based on the Yul-Walker approach as follows
\begin{equation}
\p_\wl(f) = \frac{\hat{\sigma}_{w}^2}{|1+\sum_{\u=1}^{\po}\hat{a}_{\u} e^{-j2\pi f \u}|^2}.
\end{equation}
In the Yule-Walker method, we hypothesize that the signal, which its power spectrum is desired (i.e., $\msrV_\wl\n$), is the output of a linear time-invariant (LTI) system where the input to the system is a zero-mean white noise. In this case, $\hat{a}_{\u}$, for ($1 \leq \u \leq \po$), is the estimated autoregressive (AR) parameter of the output signal using the Levinson Durbin algorithm, and $\hat{\sigma}_{w}^2$ is the $\po^{\text{th}}$-order minimum mean-square estimate (MMSE) of the variance of the zero-mean white noise provided as the input to the system which is computed as follows
\begin{equation}
\hat{\sigma}_{w}^2 =  \gamma_{\wl}(0)\prod_{\tilde{a}u=1}^{p}[1-|\hat{a}_{\tilde{a}u}|^2],
\end{equation}
where the biased autocorrelation estimate $\gamma_{L}(0)$ is given by
\begin{equation}
\gamma_{L}(0) = \frac{1}{\wl} \msrV^H_\wl(n) \msrV_\wl(n).
\end{equation}
 Superscript $H$ denotes hermitian transpose operator. Fig.~\ref{PSD} illustrates one sample of the power spectrum of a hand motion signal which consists of both tremor and voluntary motions. As can be seen from Fig.~\ref{PSD} and as shown in relevant literature (such as in ~\cite{veluvolu2007bandlimited, veluvolu2010estimation, Atashzar:2016}), the frequency band of the tremor is completely distinct from that of the voluntary movement. This distinction results in a valley (between the frequency contents of tremor and that of the voluntary movements) in the power spectrum of the signal. In this paper, the central frequency of the valley that separates the frequency contents of the tremor from that of the voluntary motion is called as the \textit{cut-off frequency} $(\fd)$.

The cut-off frequency $\fd$ can be used to form and extract a ground truth for the tremor signal denoted by $\msrV_{i_{\wl}}\gt\n$. For this purpose, typically~\cite{veluvolu2007bandlimited, veluvolu2010estimation,Atashzar:2016},
a sharp high-pass filtering is applied on the windowed version $\msrV_\wl\n$ of the signal with cut-off frequency $\fd$.
{\color{black} The sharp high-pass filtering approach is described in Algorithm~\ref{sharpfilter}. }
\begin{algorithm}[t!]
{\color{black}	\caption{\textproc{Sharp high-pass filtering methodology according to~\cite{Atashzar:2016}}}
	\label{sharpfilter}
	\begin{algorithmic}[1]
		\Input The measurement signal: $\msrV_\wl\n$.
		\Output Hyper-Parameters: The ground truth for the tremor signal.
		\vspace{.025in}
		\item[S1.] Calculate the Fast Fourier Transform (FFT), denoted by $\bm{F}(\msrV_\wl\n)$, of the input $\msrV_\wl\n$.
		\item[S2.] Construct an indexing vector with the same length as of $\bm{F}(\msrV_\wl\n)$, i.e., the index vector has ones for frequencies greater than $\fd$ and zeros for the rest.
		\item[S3.] Multiply the indexing matrix by $\bm{F}(\msrV_\wl\n)$ in an element-wise fashion; $\bigg[ \bm{F}(\msrV_\wl\n)\bigg]_{indexed}$.
		\item[S4.] Apply inverse-FFT to $\bigg[ \bm{F}(\msrV_\wl\n)\bigg]_{indexed}$.
	\end{algorithmic}}
\end{algorithm}
However, in the existing methodologies a pre-defined and fixed value of $\fd$ is used to extract the ground truth. In contrary, we propose to incorporate an autonomous tremor frequency band detection algorithm by capitalizing on the smoothness of the curve of the power spectrum density, and the apparent distinction between the frequency contents of tremor versus that of the voluntary motion. The proposed autonomous detection algorithm computes the minimum value of the valley in the power spectrum density curve which is a representative of the cut-off frequency $\fd$. Thus, the cut-off frequency $\fd$ is not a pre-defined or fixed value, instead, its estimated value is updated in a real-time fashion at each iteration of the $\HPA$ scheme.

After high-pass filtering the measurement signal, and extracting the ground truth for the tremor signal $\msrV_{i_{\wl}}\gt\n$, we derive the ground truth for voluntary movement  as follows
\begin{eqnarray}
\msrV_{v_{\wl}}\gt\n = \msrV_\wl\n - \msrV_{i_{\wl}}\gt\n.
\end{eqnarray}
The final step of the $\HPA$ scheme is to find the optimum values for the adaptive weights denoted by $\w_\wl\n$ (i.e., Hyper-parameters) for the current iteration which is then used by the $\RTP$ scheme to fuse wavelet approximations and construct the voluntary movement.
In other words, the $\w_\wl\n$ vector defines a weight for each approximation in a way that the weighted mixture of the approximations results in an optimized reconstructed voluntary signal.
To achieve this goal, a wavelet decomposition is applied to the windowed signal $\msrV_\wl\n$ for extracting $\ld$ number of approximations. The windowed signal $\msrV_\wl\n$ can then be represented by a combined series expansion using both the approximation coefficients and the wavelet (detail) coefficients as follows
\begin{equation}
\msrV_\wl\n = \sum_{\u=-\infty}^{\infty}\S(J,\u)\phi_{J,\u}(n) +  \sum_{j=1}^{J} \sum_{\u=-\infty}^{\infty}\T(j,\u)\psi_{j,\u}(n).\label{eq:Recons}
\end{equation}
In order to reconstruct the $j^{\text{th}}$ approximation of the signal $\a_{j}$, for ($1 \leq j \leq \ld$), according to the multilevel behavior of the wavelet transforms, we need to incorporate the detail and approximation coefficients of the signal from the ($j+1$)$^{\text{th}}$ level of decomposition. On the other hand, as the DWT requires having a discrete time signal as an input which has fixed sampling rate, we extract the approximation coefficients~as
\begin{equation}\label{disappcoeff}
\S(j,\u) = \sum_{n=1}^{\wl} \msrV_\wl\n\phi_{j,\u}\n,
\end{equation}
and the detail coefficients as in
\begin{equation}
	\T(j,\u) = \sum_{n=1}^{\wl} \msrV_\wl\n\psi_{j,\u}(n).
\end{equation}
The approximation signal in level $j$ (i.e., $\a_{j}$), for ($1 \leq j \leq \ld$), can be reconstructed as follows,
\begin{equation}\label{apprecuns}
\a_{j}\n = \sum_{\u=-\infty}^{\infty}\S(j+1,\u)\phi_{j+1,\u}\n + \sum_{\u=1}^{\infty} \T(j+1,\u)\psi_{j+1,\u}(n).
\end{equation}
Reconstructed approximations at all levels are represented in compact form as matrix $\A_\wl\n = \{\a\j\}_{1 \leq j \leq \ld} \in \mathbb{R}^{\wl\times\ld}$.
We note that extracted approximation coefficients together with the detailed coefficients form the original signal (Eq.~\eqref{eq:Recons}), while the reconstructed approximations provide a smoothed version of the signal without the need for the detail coefficients. The levels of decomposition and the mother wavelet function which is deployed are discussed later in Section~\ref{sec:exp}.

Once the reconstructed approximation matrix $\A_\wl\n$ is computed, an unconstrained nonlinear optimization based on Quasi-Newton technique~\cite{Rioul:1991} is deployed to derive the adaptive weight vector $\w_\wl\n \in \mathbb{R}^{J}$ such that the  error $e_\wl\n$ defined below is minimized
\begin{eqnarray}\label{optimizeerror}
e_\wl\n =
\sqrt{\frac{1}{L}\big(\A_\wl\n\w_\wl\n-\msrV_{v_{\wl}}\gt\n\big)^T\big(\A_\wl\n\w_\wl\n
-\msrV_{v_{\wl}}\gt\n\big)}.
\end{eqnarray}
This completes one iteration of the $\HPA$ scheme. Vector $\w_\wl\n$, the cut-off frequency $\fd$, and the ground truth values $\msrV_{v_{\wl}}\gt\n$ are the Hyper-parameters constituting output of the $\HPA$~scheme and are fed into the $\RTP$~scheme described next.  Algorithm~\ref{Alg1} summarizes the $\HPA$~scheme.

%---------------------------------------------------------------------
\begin{algorithm}[t!]
\caption{\textproc{The $\HPA$ Framework}}
\label{Alg1}
\begin{algorithmic}[1]
\Input The measurement signal: $\msrV_\wl\n$.
\Output Hyper-Parameters: Optimized weight vector $\w_\wl\n$; Discriminant frequency $f_d$; and Estimated ground truth vector $\msrV_{v_{\wl}}\gt\n$.
\vspace{.025in}
\item[S1.] Compute the Yul-Walker power spectrum density of $\msrV_\wl\n$; $\p_L(f)$.
\item[S2.] Calculate the corresponding frequency to the minimum of valley in $\p_L(f)$; $f_d$.
\item[S3.] Decompose $\msrV_\wl\n$ to \textit{J} levels of approximation and form matrix $\A_\wl\n$.
\item[S4.] Derive the optimized matrix $\w_\wl\n$ according to Eq.~\eqref{optimizeerror}
\end{algorithmic}
\end{algorithm}
%---------------------------------------------------------------------
%=========================================================
\subsection{Real-Time Tremor Prediction (\RTP) Scheme}
%=========================================================
The $\RTP$~scheme performs real-time data analysis on the measurement signal $\msrV\n$ at the current time instant $n$, and provides as an output  a myopic predicted version of the tremor signal for the next time instant ($n+1$). In this scheme, some predefined parameters are fixed and do not change over time, but some of the hyper-parameters are updated via the $\HPA$~scheme. The $\HPA$~scheme provides optimized hyper-parameters for the $\RTP$~to increase the overall real-time performance of the system for tremor extraction. In this scheme, the KF and wavelet decompositions are jointly incorporated to provide predictions for the tremor signal.

Since we are deploying the wavelet decompositions in this scheme and we are also aiming at a real-time system, it is important to keep the amount of calculations as low as possible. The minimum length of a signal which is going to be decomposed with wavelet transforms, should be at least $2^{\ld}$. Hence, the $\RTP$~scheme operates on the last $2^\ld$ samples of the measurement signal $\msrV_\ld\n$ defined as follows
\begin{eqnarray}
\msrV_\ld\n  \triangleq [\msr(n-2^\ld+1), \ldots,  \msr\n]^T.
\end{eqnarray}
Now, we need to incorporate  matrix $\A_\ld$ in the KF formulations. Please note that for each time sample that the system receives one new measurement and $n$ increases, the \RTP~scheme slides one sample ahead, and a new $\msrV_\ld\n$ is formed. We also define the independent variable $k$ to demonstrate the calculations within the \RTP~scheme.
At the first step, the signal $\msrV_J\n$ is decomposed into $J$ levels and $J$ approximation signals are extracted. Then the approximations are put together to form the matrix $\A_\ld \in \mathbb{R}^{ J\times 2^J }$. Note that the number of columns in matrix $\A_\ld$ (computed within the $\RTP$~scheme) is different from that of matrix $\A_\wl$ that is computed within the $\HPA$~scheme.

We propose to use Kalman filter to recursively iterate over the last $2^\ld$ samples to compute a predictive value for the approximations in the next coming iteration. For example, when the current iteration is $k$, samples ($n-2^\ld+1$) to ($n$) are used to provide predictions for time iteration $n+1$. Index $k$ is chosen to discriminate between the independent variable $n$ that is used for the whole time domain of the problem. Note that $k$ ranges from $1$ to $2^\ld$.
Within the framework of the KF, we define columns $\a_\ld(\cdot)$ of matrix $\A_\ld$ as the state vector, i.e.,
\begin{eqnarray}\label{eq:KFState}
\x\k \triangleq \a_\ld\k.
\end{eqnarray}
The following linear state-space model is considered to represent the evolution of state vector over time
\begin{eqnarray}\label{differenceRTP}
\x\k= \F\x\pk + \w\k,
\end{eqnarray}
where matrix $\F$ is a diagonal matrix $\bm{I}_\ld$ of appropriate dimension. Observed signal values are considered as the measurement $z\k$ in the KF recursions, i.e.,
\begin{eqnarray}
z\k \triangleq m_\ld\k.
\end{eqnarray}
The observation model is constructed based on the hyper-parameters (vectors $\w_\wl\n$ and $\msrV_{v_{\wl}}\gt\n$) which are provided by the $\HPA$ scheme. The observation model is set to the optimized weights $\w_\wl\n$ reported by the $\HPA$~scheme. It is worth mentioning that the weight vector $\w_\wl\n$ is updated whenever the $\HPA$ scheme re-runs; however, the KF matched to that window uses the same model during its operation. Therefore, index $k$ is not required here.
The observation noise $v\k$ is considered to be a zero mean Gaussian distribution with variance $R$ which is equal to the variance of the ground truth for tremor signal. The ground truth for tremor signal ($\msrV_{v_{\wl}}\gt\n$) is considered as the known bias of the model. The observation model is, therefore, given by
\begin{eqnarray}\label{measureRTP}
z\k = \w_\wl\n^T \x\k + m_{v_{\wl}}\gt\k + v\k.
\end{eqnarray}
In this context, the Kalman Filter recursions are used to provide myopic predictions as follows
\begin{eqnarray}
\hx\kpk &=& \F\hx\pkpk \\
\P\kpk &=& \F\P\pkpk\F^{H}+ \Q.
\end{eqnarray}
By incorporation of new observation, the states are updated~as
\begin{eqnarray}
\br\k &=& \z\k -  \w_\wl\n^T \hx\kpk - \msr_{v_{\wl}}\gt\k \label{eq:Innov} \\
\S\k &=& \w_\wl\n^T\P\kpk \w_\wl\n+ R. \label{eq:inovCov}\\
\K\k &=& \P\kpk\w_\wl\n\S\k^{-1} \label{eq:KalGain}\\
\hx\kk &=& \hx\kpk+ \K\k \br\k \\
\P\kk &=& \P\kpk - \K\k\w_\wl^{T}\n\P\kpk.\label{eq:CovUpd}
\end{eqnarray}
Once iterations of the Kalman filter based on Eqs.~\eqref{eq:Innov}-\eqref{eq:CovUpd} is completed based on observations $z\k$, for ($1 \leq k \leq 2^J$), the last updated estimate $\hx\k$ is predicted one-step forward and $\hx\kpk$ is used as the predicted state for next time sample which has not happened yet. This myopic prediction is used to form a predicted estimate for the voluntary movement signal in the next time sample.  Algorithm~\ref{Alg2} summarizes different steps of the $\RTP$ scheme.

%---------------------------------------------------------------------
\begin{algorithm}[t!]
	\caption{\textproc{The $\RTP$ Framework}}
	\label{Alg2}
	\begin{algorithmic}[1]
		\Input The measurement signal: $\msrV\n$, and; The optimized weights; $\w_\wl\n$.
		\Output The predicted voluntary signal; $\volpr_v\n$
		\vspace{.025in}
		\item[S1.] Segment the last $J$ samples of the measurement signal and form $\msrV_J\n$
		\item[S2.] Perform KF on $\msrV_J\k$ based on Eqs.~\eqref{eq:KFState}-\eqref{eq:CovUpd} for $k=1,\cdots,2^\ld$
		\item[S3.] Compute $\volpr_{v}\n = \w^{T}_\wl\n\hx\kpk$ for $k=2^J+1$ as the prediction for the next time instance.
	\end{algorithmic}
\end{algorithm}
%---------------------------------------------------------------------

%=========================================================
\subsection{The Overall Work-flow}
%=========================================================
After describing the $\RTP$~and $\HPA$~schemes, now we outline the overall work-flow of the proposed $\WAKE$ framework. As we have discussed earlier, the $\HPA$~scheme, operates when $L$ new measurement samples are available. Hence, when the algorithm starts, the algorithm waits to receive the first $L$ samples to adjust the hyper-parameters for the $\RTP$~scheme, then the $\RTP$~scheme starts operating. The $\RTP$ continues till $L$ new samples are ready for the $\HPA$. Algorithm~\ref{Alg3} outlines the overall work-flow of the $\WAKE$ framework.

%---------------------------------------------------------------------
\begin{algorithm}[t!]
	\caption{\textproc{The Whole System Framework}}
	\label{Alg3}
	\begin{algorithmic}[1]
		\Input The measurement signal; $\msrV\n$.
		\Output The predicted voluntary signal; $\volpr_v\n$
		\vspace{.025in}
		\item[S1.] Wait until $\wl$ number of samples are ready.
		\item[S2.] Form $\msrV_\wl\n$
		\item[S3.] $[\w_\wl,~\fd]~=~\HPA(\msrV_\wl\n)$
		\Loop
		\If{$modulo(n/\wl) \in \mathbb{Z}$}
			\State $[\w_\wl,~\fd]~=~\HPA(\msrV_\wl\n)$
		\Else
			\State Execute the $\RTP$~scheme:
		
		\For {\text{$\{k=1,...,2^J\}$}}
		
			\State Form $\msrV_J\n$
			\State Perform KF on $\msrV_J\n$ based on Eqs.~\eqref{eq:KFState}-\eqref{eq:CovUpd}.
		\EndFor
		\State $\volpr_v(n+1) = \w^T_\wl\n \hx\kpk |_{k=2^J+1}$
		\EndIf
		\EndLoop
	\end{algorithmic}
\end{algorithm}
%---------------------------------------------------------------------

%OOOOOOOOOOOOOOOOOOOOOOOOOOOOOOOOOOOOOOOOOOOOOOOOOOOOOOOOO
\section{Simulation Experiments/Results}  \label{sec:exp}
%OOOOOOOOOOOOOOOOOOOOOOOOOOOOOOOOOOOOOOOOOOOOOOOOOOOOOOOOO
In this section, we present various simulation and experimental results to evaluate the performance of the proposed tremor extraction framework. First, in Sub-section~\ref{subsec:Exp1}, we evaluate effects of different design variables/parameters on the overall performance of the proposed $\WAKE$ framework. Second, in Sub-section~\ref{sec:result}, we evaluate the overall performance of the $\WAKE$ framework based on synthetic and real datasets.
For performance evaluation,  we calculate normalized root mean square error (RMSE)~\cite{Atashzar:2016} as given below
\begin{eqnarray}
\text{NRMSE}= \frac{\sqrt{\frac{1}{N}\sum_{n=1}^{N}\big({\msrV_v\gt\n-\volpr_v\n}\big)^2}}{\max(\msrV_v\gt) - \min(\msrV_v\gt)}.\label{rmse1}\label{rmse2}
\end{eqnarray}
Here, the ground truth ($\msrV_i\gt\n$) for tremor signal is obtained by off-line sharp filtering of the measurement based on the mean of $f_d$ over all of the executions of the $\HPA$ scheme. The predicted tremor ($\hat{\msrV}_i\n$) is obtained by subtracting the predicted voluntary movement ($\volpr_v\n$) from the measurement signal.

%=========================================================
\subsection{Parameter Selection/Analysis} \label{subsec:Exp1}
%=========================================================
In this sub-section, we provide the results of several experiments performed to properly select the following design variables and parameters: (i) The window length ($L$); (ii) Mother wavelet function and its effect on the overall performance; (iii) Level of decomposition in wavelet transform and its effects on the performance, and; (iv) The model-order ($p$ in the Yul-Walker power spectrum estimation). To have a fair comparison, all the experiments are performed based on the same measurement signal with its power spectral density (PSD) shown in Fig.~\ref{PSD}.

%-------------------------------------------------------------------------------------------------------------
\subsubsection{Effects of Window Length ($L$)}
\begin{figure}[t!]
\centering
\mbox{\subfigure[]{\includegraphics[width=8cm, height=6.5cm]{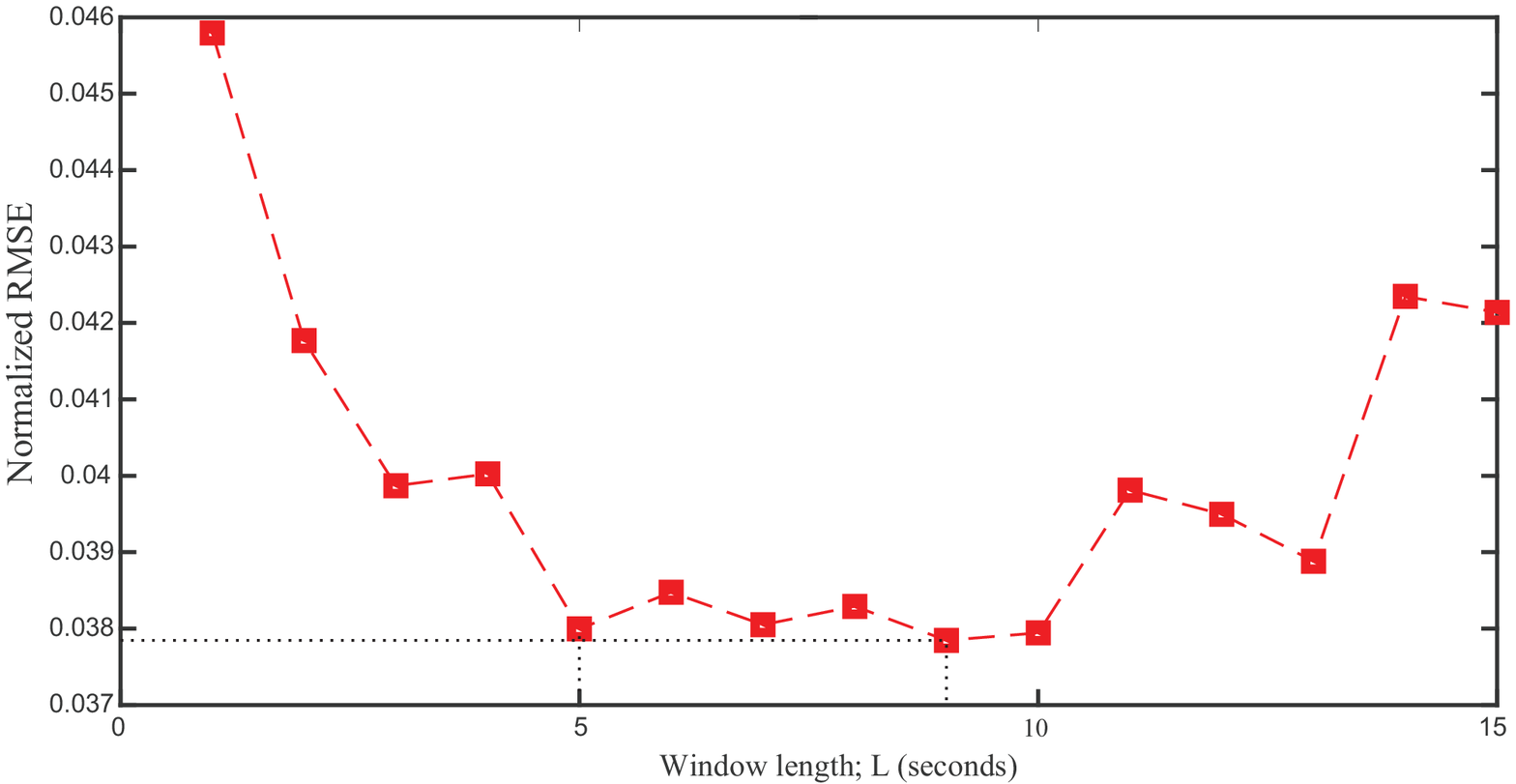}}
\subfigure[]{\includegraphics[width=8cm, height=6.5cm]{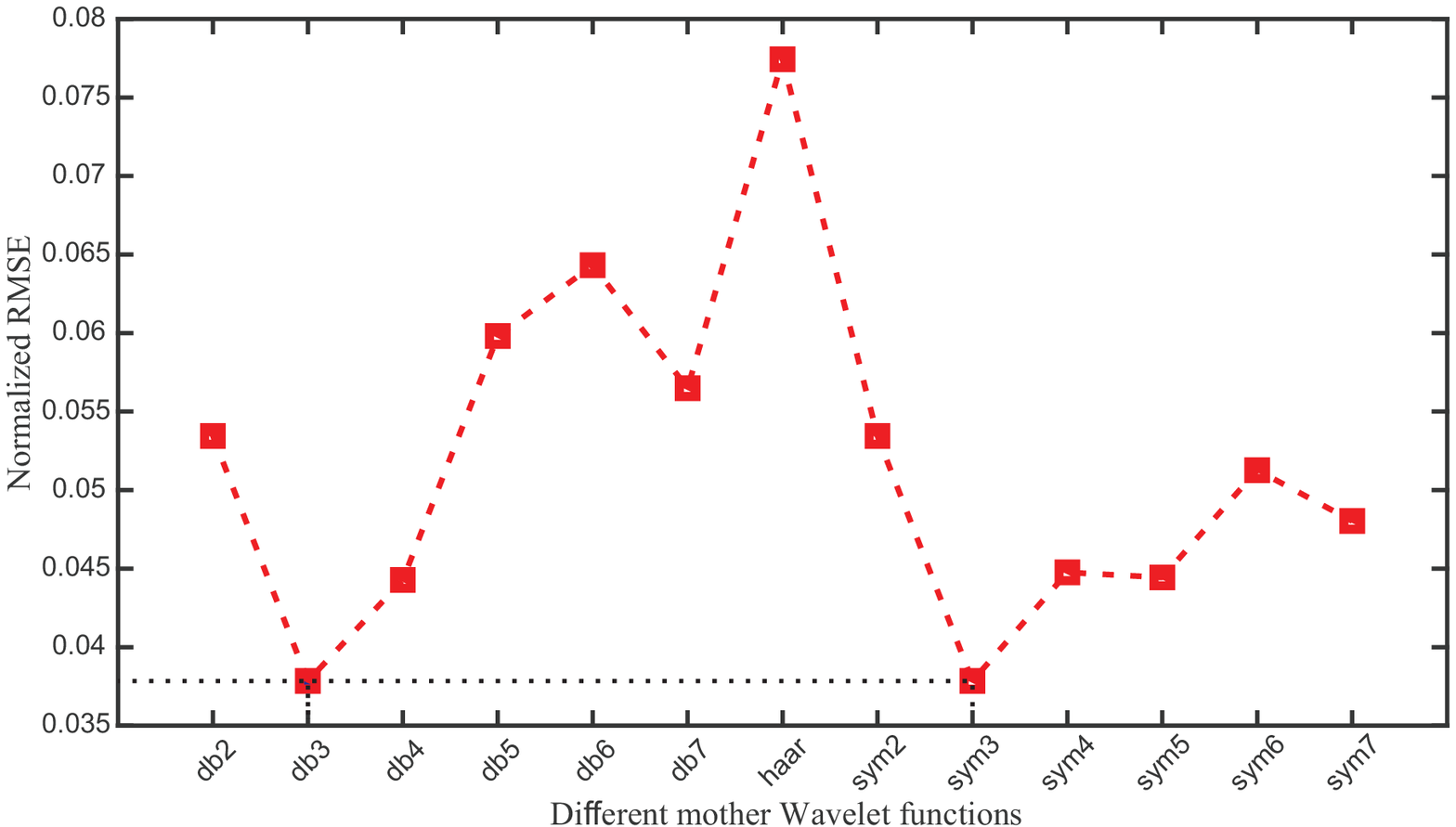}}}
\caption{(a) NRMSE computed based on varying window length, $L$. (b) Effect of different mother wavelet functions in the overall performance. \label{testL} }
\end{figure}

%-------------------------------------------------------------------------------------------------------------
In order to select the best value for the window length $L$, we fixed all the other parameters except the value for $L$ which is varying within the range of $1$-$15$~seconds with steps of $1$~second. Fig.~\ref{testL}(a) illustrates the variation of the overall performance as a function of the changes in the window length $L$. From the results, it is observed that the minimum error (maximum achievable performance) occurs at $L=9$~seconds which results in NRMSE of $0.0378$. It is worth mentioning that the performance of the system at $L=5$ seconds is rather close to that of the case with $L=9$. To construct a more dynamic system for tracking the variation in behavior of the tremor signal with less delay, we set the window length equal to $5$ seconds for the following experiments.
%------------------------------------------------------------
%------------------------------------------------------------

%-------------------------------------------------------------------------------------------------------------
\subsubsection{Effects of Mother Wavelet}
%-------------------------------------------------------------------------------------------------------------
Both the $\RTP$~and $\HPA$~schemes use the mother wavelet function, therefore, initially and at this point we assess the overall performance of the $\WAKE$ framework by investigating different mother wavelet functions including: Haar wavelet, Daubechies  (db) wavelets, and Symlets (sym) mother functions. As shown in Fig.~\ref{testL}(b), we compared the following mother functions: \textit{``db2", ``db3", ``db4", ``db5", ``db6", ``db7", ``haar", ``sym2", ``sym3", ``sym4", ``sym5", ``sym6", ``sym7"}.  It is observed that the minimum NRMSE value is obtained based on \textit{``db3"} and \textit{``sym3"} mother function, therefore, in the following experiments we mainly use the ``sym3'' wavelet.

%-------------------------------------------------------------------------------------------------------------
\subsubsection{Effect of Different Levels of Decomposition ($J$)}
\begin{figure}[t!]
\centering
\mbox{\subfigure[]{\includegraphics[width=8cm, height=6.5cm]{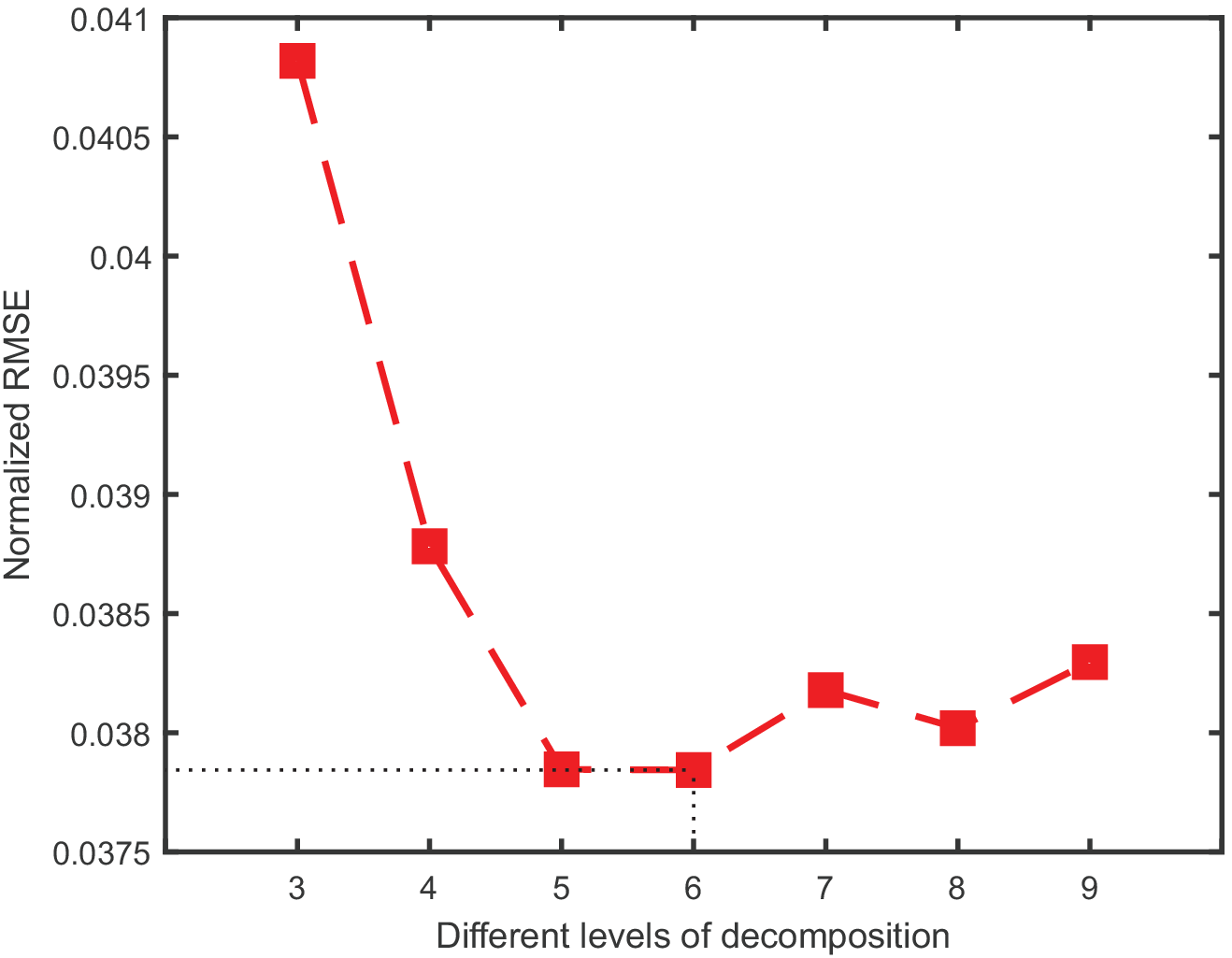}}
\subfigure[]{\includegraphics[width=8cm, height=6.5cm]{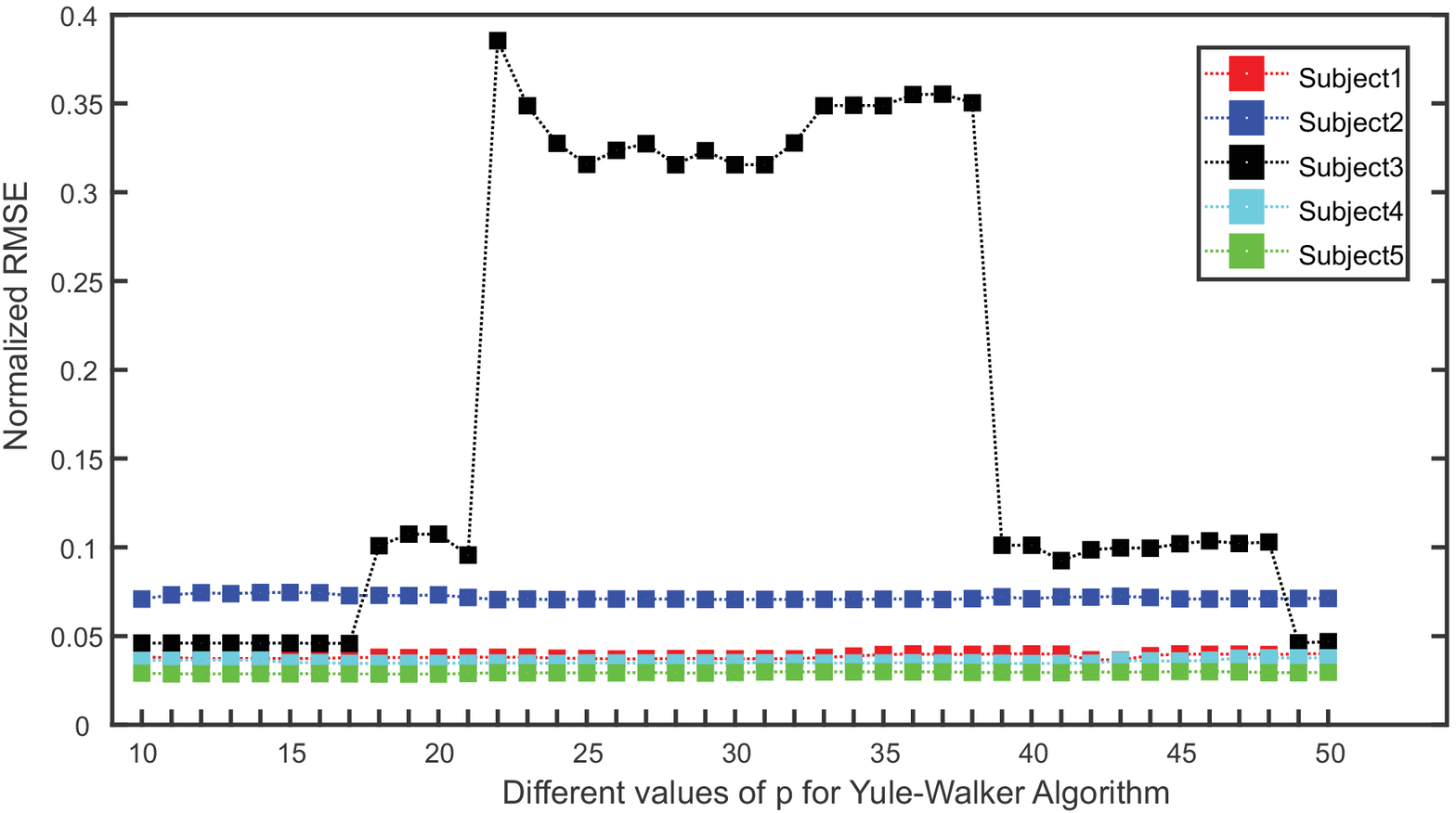}}}
\caption{(a) Effect of different levels of decomposition in wavelet transforms in the overall performance. (b) Effect of parameter $p$ in Yul-Walker method for 5 different subjects. \label{testlevels}}
\end{figure}

%-------------------------------------------------------------------------------------------------------------
Here, we investigate effects of the decomposition levels ($J$) on the overall performance in terms of the achievable NRMSE.  The level of decomposition directly influences  the length of the sliding window in the $\RTP$~scheme, since the length of sliding window is equal to $2^J$.  The results of this experiment are depicted in Fig.~\ref{testlevels}(a) where it is observed that the minimum NRMSE of  $0.03784$ is achieved with $J = 6$. We decided to set $J =5$ as the NRMSE associated with $J=5$ is $0.03785$, which is fairly close to the minimum value while  $J = 5$ levels of decomposition imposes less computational burden.
%------------------------------------------------------------
%------------------------------------------------------------
%
\begin{table}	
	\footnotesize
	\caption{Results of the BMFLC, E-BMFLC and $\WAKE$ frameworks on the synthetic data. 	\label{results2}}
	\begin{tabular}{|c|c|c|c|c|c|c|c|c|c|c|}
		\hline
		\!\!Signal  \!\!&\!\! $SNR_{in}$ \!\!& \multicolumn{3}{|c|}{PRF\%} & \multicolumn{3}{|c|}{SNR$_{out}$} & \multicolumn{3}{|c|}{NRMSE}\\
		\cline{3-11}
		& & BMFLC & E-BMFLC & \WAKE & BMFLC & E-BMFLC & \WAKE & BMFLC & E-BMFLC & \WAKE \\		\hline
		\!\!Signal1\!\! &\!\! 7.5 \!\!&\!\! 11.97 & 12.84 &\textbf{4.3} & 9.22 & 8.91 &\textbf{13.67}&0.12 & 0.13&\textbf{0.0734} \\
		\hline
		\!\!Signal2\!\! &\!\!5.1\!\!&\!\!15.94&16.41&\textbf{7.42}&\textbf{7.98}&7.85&7.42&0.14&0.14&\textbf{0.0964} \\
		\hline
		\!\!Signal3\!\! &\!\!0.3\!\!&\!\!31.49&30.72&\textbf{22.95}&5.02&5.13&\textbf{6.39}&\textbf{0.16}&0.2&0.1695 \\
		\hline
		\!\!Signal4\!\! &\!\!-4\!\!&\!\!74.2&\textbf{67.89}&68.8&1.3&1.68&\textbf{1.72}&0.3&\textbf{0.29}&\textbf{0.2904} \\
		\hline
		\!\!Signal5\!\! &\!\!4.6\!\!&\!\!22.23&26.23&\textbf{8.13}&6.53&5.81&\textbf{10.9}&0.17&0.18&\textbf{0.1009} \\
		\hline
		\!\!Signal6\!\! &\!\!0.3\!\!&\!\!57.45&73.29&\textbf{45.56}&2.41&1.35&\textbf{3.32}&0.27&0.3&\textbf{0.2414} \\
		\hline
		\!\!Signal7\!\! &\!\!-4.9\!\!&\!\!183.18&240.61&\textbf{135.74}&-2.63&-3.81&\textbf{-1.23}&0.48&0.55&\textbf{0.4077} \\
		\hline
		\!\!Signal8\!\! &\!\!20.9\!\!&\!\!4.83&4.15&\textbf{0.32}&13.16&13.82&\textbf{25}&0.08&0.07&\textbf{0.02} \\
		\hline
		\!\!Signal9\!\! &\!\!15.1\!\!&\!\!5.43&4.66&\textbf{0.9}&12.65&13.32&\textbf{20.46}&0.08&0.08&\textbf{0.0335} \\
		\hline
		\!\!Signal10\!\! &\!\!10\!\!&\!\!7.36&6.25&\textbf{2.71}&11.33&12.04&\textbf{15.67}&0.1&0.09&\textbf{0.0583} \\
		\hline		
	\end{tabular}
\end{table}
%
%-------------------------------------------------------------------------------------------------------------
\subsubsection{Effect of Model-order ($p$)}
%-------------------------------------------------------------------------------------------------------------
The model-order ($p$) is the order of the $AR$ model which is fitted to the data, therefore, its variation has direct impact on the frequency band selection which in turn would impact the overall performance of the $\WAKE$ framework. Based on our experiments, we observed that effect of the model-order is subject dependent and in general it does not follow any particular pattern. Fig.~\ref{testlevels}(b) shows the NRMSE results as a function of the model-order based on 5 different subjects. By considering the fact that increasing the value of $p$ will increase the amount of processing workload, we selected $p = 17$ in order to maintain the generality of the algorithm for different subjects.

%=========================================================
\subsection{Experiments/Results}  \label{sec:result}
%=========================================================
\textcolor{black}{In this sub-section, we evaluate the adaptive auto-adjustable $\WAKE$  framework based on three different datasets: (i) A synthetic dataset (Section~\ref{sec:syntData}); (ii) A real-dataset containing recordings from patients with action tremor (Section~\ref{sec:DataSet2}), and; (iii) A second set of real data collected from patients with Parkinson's disease (Section~\ref{subsec:spainset})}. All the design variables and parameters are selected based on the results of the experiments discussed above.
The results of the proposed method are compared with the results of two successful techniques in tremor estimation and extraction which are \textit{BMFLC}~\cite{veluvolu2007bandlimited,veluvolu2010estimation} and \textit{E-BMFLC}~\cite{Atashzar:2016}.
%-------------------------------------------------------------------------------------------------------------
\subsubsection{Synthetic Dataset}\label{sec:syntData}
%-------------------------------------------------------------------------------------------------------------
Initially, we evaluate the performance of the $\WAKE$~framework in comparison to its state-of-the-art counterparts based on a designed synthetic dataset. The introduced synthetic dataset is critically important and provides priceless insights as it allows us to precisely track and assess different aspects of the $\WAKE$ framework and observe the effects of different features.
Although synthetic datasets may not fully represent the original phenomena, we designed it in a way to carry on the following two main features of PHT: (i) In most cases, there is a clear distinction between the frequency band of voluntary movement and that of the tremor, and; (ii) Tremor happens in a certain and limited frequency band.
The synthetic signal is constructed by combining three different signals as $\s(n)=\s_v(n)+\s_w(n)+\s_t(n)$, where $\s_v$ is a sinusoidal  with a dominant amplitude (compared to the other two signals) with frequency of less than $1$Hz representing the voluntary movement.  The second signal $\s_w$  is an additive white noise occurring at all frequencies and is generated by setting a frequency resolution (frequency gap) and then adding sinusoidal with different frequencies between 0 to $F_s/2$ with steps equal to the frequency resolution. The amplitude of sinusoidal follow a uniform distribution with specified mean and variance.  The third signal $\s_t(n)$ represents the tremor and its frequency band is limited to $6$-$14$Hz to be consistent with the band limits specified in~\cite{veluvolu2010estimation}. The tremor signal $\s_t(n)$ is generated following the same procedure used to generate the white noise. Finally, for both $\s_t(n)$ and $\s_w(n)$, a random phase is added to the constituting sinusoidal in order to prevent synchronization of the sinuses resulting in formation of ``Sinc'' waveforms.
Based on the above procedure, ten test signals are constructed sampled at $F_s = 100$Hz and each representing $50$ seconds of activity. Each of the ten constructed signals represents a different signal to noise ratio (SNR) calculated as
\begin{eqnarray}
\text{SNR} = 10 \times \log_{10}(\frac{P_{sig}}{P_{noise}}).
\end{eqnarray}
We note that in our calculations, we use $P_{noise} = P(s_w) + P(s_t)$ and $P_{sig} = P(s_v)$, where $P(\cdot)$ denotes the power of a signal. In this evaluation, beside the NRMSE and SNR, we also use the power ratio factor (PRF) defined as follows
\begin{eqnarray}
	\text{PRF}(n) = \frac{(\s_v(n) - \volpr_v(n))^2}{(\s_v(n))^2} \times 100,
\end{eqnarray}
which is used as an alternative measure to evaluate the performance of the proposed  $\WAKE$~framework. The results are demonstrated in Table.~\ref{results2}. It is observed that the $\WAKE$~significantly outperforms its counterparts.

\textcolor{black}{As a final note, we would like to point out that unlike the currently existing techniques, the proposed $\WAKE$ framework demonstrates a deterministic processing pipeline, which does not require any parameter and structure tuning based on the input data. The technique can even automatically finds the frequency range of interest. This significantly enhances the practicality of the proposed approach specially for clinical and biomedical settings, under which excessive parameter tuning and sensitivity to the choice of parameters are not acceptable. As a result, it can be mentioned that the proposed technique is not a data-driven processing framework which requires several training examples to fine-tune the parameters. Thus, a higher number of data points, would not provide a better model and it would not change the performance of the technique.  However, higher number of data points can help to better judge the performance, as such we are reporting below the evaluation results based on two real datasets in addition to the synthetic dataset considered here. In short, the results of all evaluations supports the benefit of the proposed framework. For the evaluation phase, we believe that the synthetic data used in this sub-section provides a robust assessment tool for evaluating the proposed \WAKE~framework. Synthetic dataset helps us to precisely track the changes and transformations applied to the three components of the synthetic signal which are the voluntary signal, the involuntary signal and the white noise. Thanks to this dataset, we can more reliably calculate the signal-to-noise ratio, and also it enables us to understand how differently the $\WAKE$~algorithm handles the involuntary movement and the white noise.}

%-------------------------------------------------------------------------------------------------------------
\subsubsection{The Motus Dataset}\label{sec:DataSet2}
%-------------------------------------------------------------------------------------------------------------
{\color{black} The Motus dataset is recorded with bi-axial gyroscopes which are developed at Motus Bioengineering Incorporation, Benicia, CA and is publicly available~\cite{reeke2005modeling}. The collected signals associated with different subjects with hand tremor are recorded with sampling frequency of $100$ Hz. The subjects with action tremor are asked to perform pronation/supination with their hand, while the gyroscope, mounted on their hand, is recording the movement's signal. The recordings, therefore, contain information about the voluntary movement (pronation/supination), as well as information corresponding to the  action tremor. The results of applying the proposed $\WAKE$~on the Motus dataset are presented in Table~\ref{results1}. Samples of the tremor signal and the extracted voluntary component are shown in Figs.~\ref{motus1} and \ref{motus4}. Similarly, a sample of the measurement signal and the extracted voluntary component is illustrated in Fig.~\ref{motusMeasFig}. As it is observed from the results, beside the fact that the $\WAKE$ framework provides real-time tremor extraction/estimation, it provides drastically improved results and significantly outperforms its well regarded counterparts.

As a side note to our discussion, we would like to also elaborate on better performance of the E-BMFLC in comparison to the BMFLC.  In brief, the harmonic model used in the BMFLC is much smaller and different from the model used in the E-BMFLC. This means that in the BMFLC technique, we only have the harmonic model for the frequency range of tremor and just the harmonics for tremor are estimated and updated. In other words, the BMFLC does not utilize a general large harmonic model that models the whole signal. Instead, it tries to keep track of the pre-defined frequency band of the tremor. For example, for a tremor occurring in the range of $12$ to $20$ Hz, the BMFLC fits a model of harmonics from $12$ and $20$ Hz. On the contrary, the E-BMFLC models the range from $0$ to $20$ Hz and after estimating the parameters in each iteration, a part of the estimated harmonics will be considered as the tremor. Consequently, the E-BMFLC method is expected to better track the dynamics of the tremor in comparison to the BMFLC.

\begin{figure}[t!]
	\centering
	\includegraphics[scale=0.7]{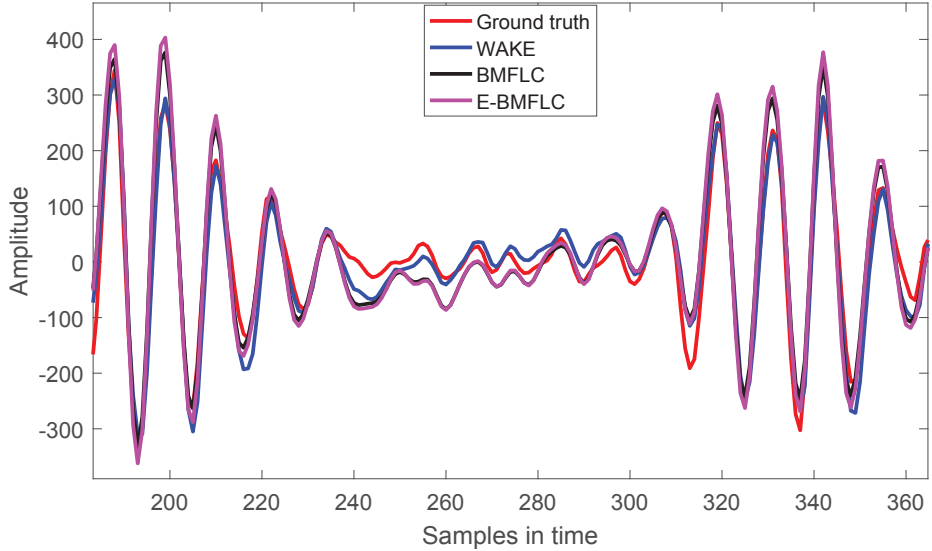}
	\caption{The predicted tremor derived by the proposed $\WAKE$ framework, the BMFLC method and the E-BMFLC method compared to the ground truth of the tremor derived by off-line processing of the signal (Subject 1 in Motus Dataset). \label{motus1}}
\end{figure}

\begin{figure}[t!]
	\centering
	\includegraphics[scale=0.7]{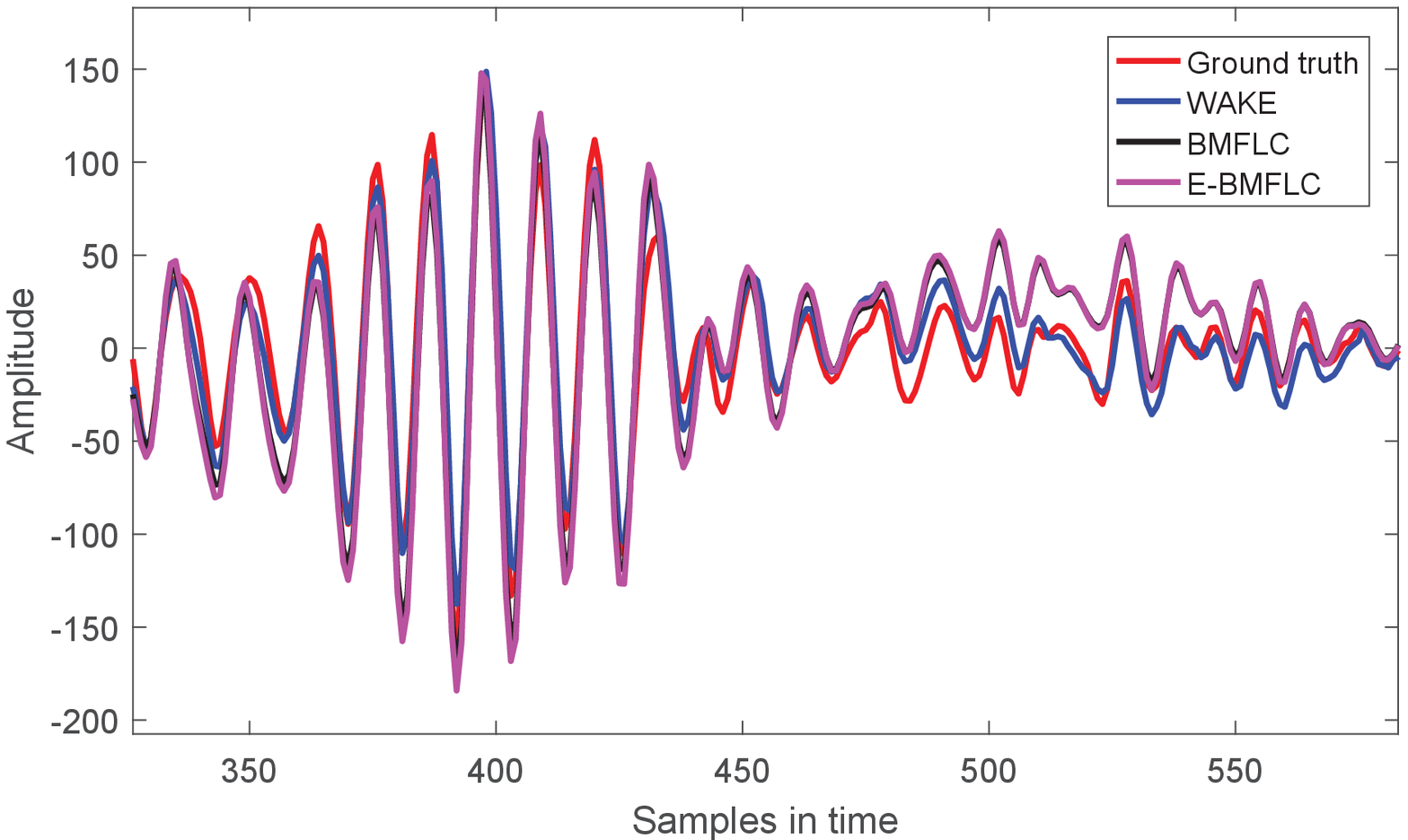}
	\caption{The predicted tremor derived by the proposed $\WAKE$ framework, the BMFLC method and the E-BMFLC method compared to the ground truth of the tremor derived by off-line processing of the signal (Subject 4 in Motus Dataset). \label{motus4}}
\end{figure}

\begin{figure}[t!]
	\centering
	\includegraphics[scale=0.5]{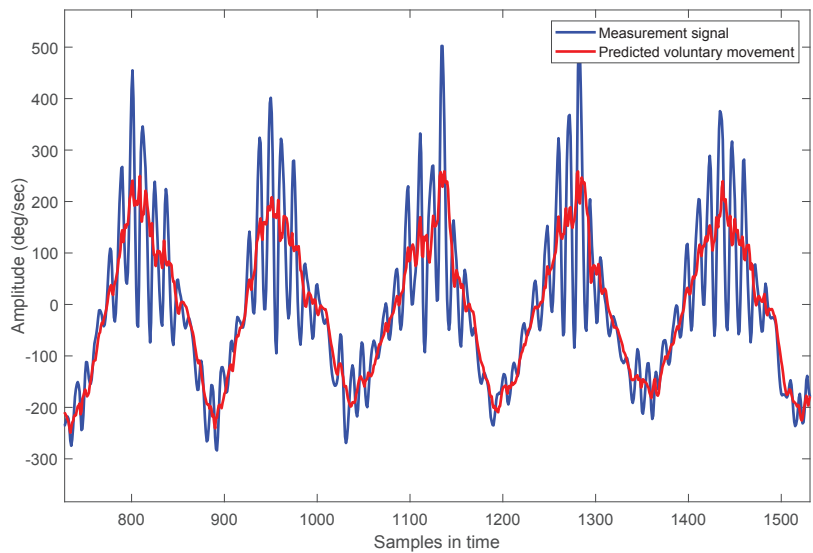}
	\caption{Visualization of the raw measurement signal and its extracted voluntary component. \label{motusMeasFig}}
\end{figure}

}
%------------------------------------------------------------
 \begin{table}[t!]
 	\centering
 	\caption{Results of the application of the proposed method for Motus Dataset. The measure of error is Normalized-RMSE.}
 	\begin{tabular}{|c|c|c|c|}
 		\hline
 		Subjects & BMFLC & E-BMFLC & $\WAKE$ \\
 		\hline
 		subject1 & 0.1130 & 0.0634 & 0.0385 \\
 		subject2 & 0.1015 & 0.1263 & 0.0721 \\
 		subject3 & 0.0927 & 0.0742 & 0.0444 \\
 		subject4 & 0.1113 & 0.0622 & 0.0355 \\
 		subject5 & 0.1043 & 0.0444 & 0.0279 \\
 		\hline
 		\textbf{Mean} & \textbf{0.1046} & \textbf{0.0741} & \textbf{0.0437} \\
 		\hline
 	\end{tabular}
  \label{results1}
   \end{table}
%------------------------------------------------------------
{\color{black}
%-------------------------------------------------------------------------------------------------------------
\subsubsection{The Smartphone Dataset}\label{subsec:spainset}
%-------------------------------------------------------------------------------------------------------------
To further elaborate on the generality and superior performance of the proposed $\WAKE$~framework, we have employed a third set of tremor recordings from 10 patients with Parkinson's Disease. This dataset is provided by ``\textit{Hospital Clinic de Barcelona, Barcelona, Spain}''~\cite{barrantes2017differential, rojas2018time}.
The signals are recorded with the built-in tri-axial accelerometer of a smartphone (in this set iPhone 5S, Apple Inc., USA). The smartphone is mounted on the dorsum of the hand which shows more sever tremor. The signals in this dataset are recorded with the sampling frequency of $100$ Hz, constituting of three channels of data, representing the acceleration in $X$-axis, $Y$-axis and $Z$-axis. The patients were asked to have their arms stretched while their upper limb is fully extended in front of them and their palms are facing the ground. Figs.~\ref{spain4} and~\ref{spain9} provide demonstrating examples of the performance of the $\WAKE$ framework in extracting the tremor signal from the Smartphone dataset.
This data set which is collected by a different device and a different team is used to illustrate the efficacy of the proposed $\WAKE$ framework in comparison to the state-of-the-art techniques.  As the measurement device for this dataset is easily accessible, this set is considered as a valuable development testbed. From clinical stand point, any diagnostic framework developed on this dataset would be practically more applicable as it does not require expensive and complex sensors. On the other hand and from signal processing stand point, the main difference between this dataset and the Motus one (utilized in Section~\ref{sec:DataSet2}) is the quality of the recordings. The Motus set is recorded with medical-grade sensors, specialized for tremor measurements, therefore, provides very clean and noise-free signals. The smartphone dataset, however, is recorded with the internal accelerometers of a smartphone, which are employed for general applications and thus the quality of the recorded signals is not as high, imposing a high degree of complexity to the processing pipeline to extract the voluntary movement. Consequently, the Smartphone dataset incorporated for performance evaluation can be considered as a difficult case study from signal processing point of view.

Table~\ref{resultsSpain} provides the performance results obtained via the proposed $\WAKE$ framework, the BMFLC, and the E-BMFLC frameworks. It is worth mentioning that the same parameter selection process is performed for the smartphone dataset, separately for each channel of data. Hence, the information in \textit{italic} font presented in Table~\ref{resultsSpain} denote the best hyper-parameters for each axis. For parameter-selection, the signals from the first subject are analyzed and then the parameters are employed for other subjects. The numbers in the table represent the NRMSE  between the predicted tremor and the ground truth signal for tremor. Figs.~\ref{spain4} and~\ref{spain9} illustrate the performance of the proposed $\WAKE$ framework compared to its counterparts including the BMFLC method and the E-BMFLC method. It is observed that the results obtained from the proposed framework are significantly improved in comparison to its counterparts.  As stated previously,  the smartphone dataset is a difficult case study from signal processing point of view, being capable of significantly outperforming the state-of-the-art on this dataset illustrates superiority of the proposed $\WAKE$ framework. This evaluation clearly shows the superior performance of the proposed $\WAKE$ approach and supports its practicality and reproducibility for signals collected with different machines.

\begin{figure}[t!]
	\centering
	\includegraphics[scale=0.7]{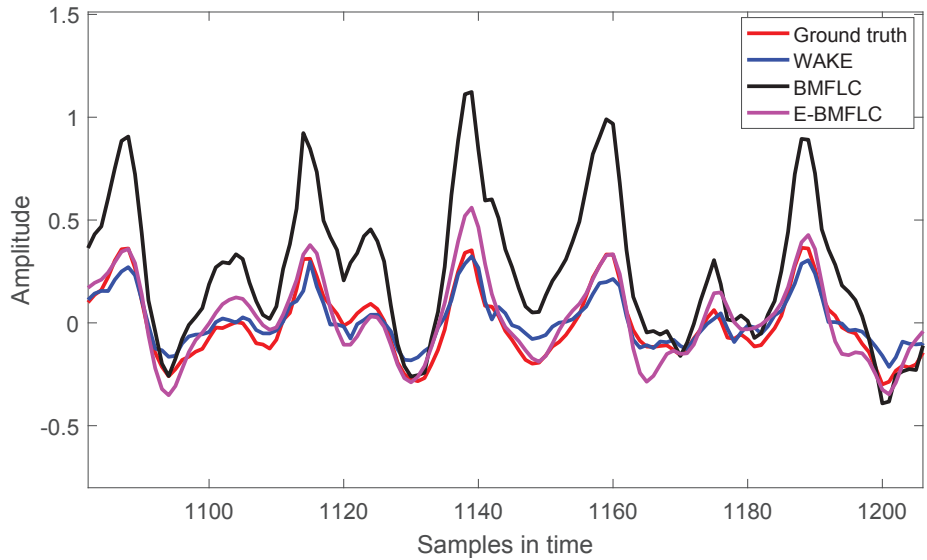}
	\caption{The predicted tremor derived by the proposed $\WAKE$ framework, the BMFLC method and the E-BMFLC method compared to the ground truth of the tremor derived by off-line processing of the signal (Subject 1 - $X$-axis in Smartphone Dataset). \label{spain4}}
\end{figure}

\begin{figure}[t!]
	\centering
	\includegraphics[scale=0.7]{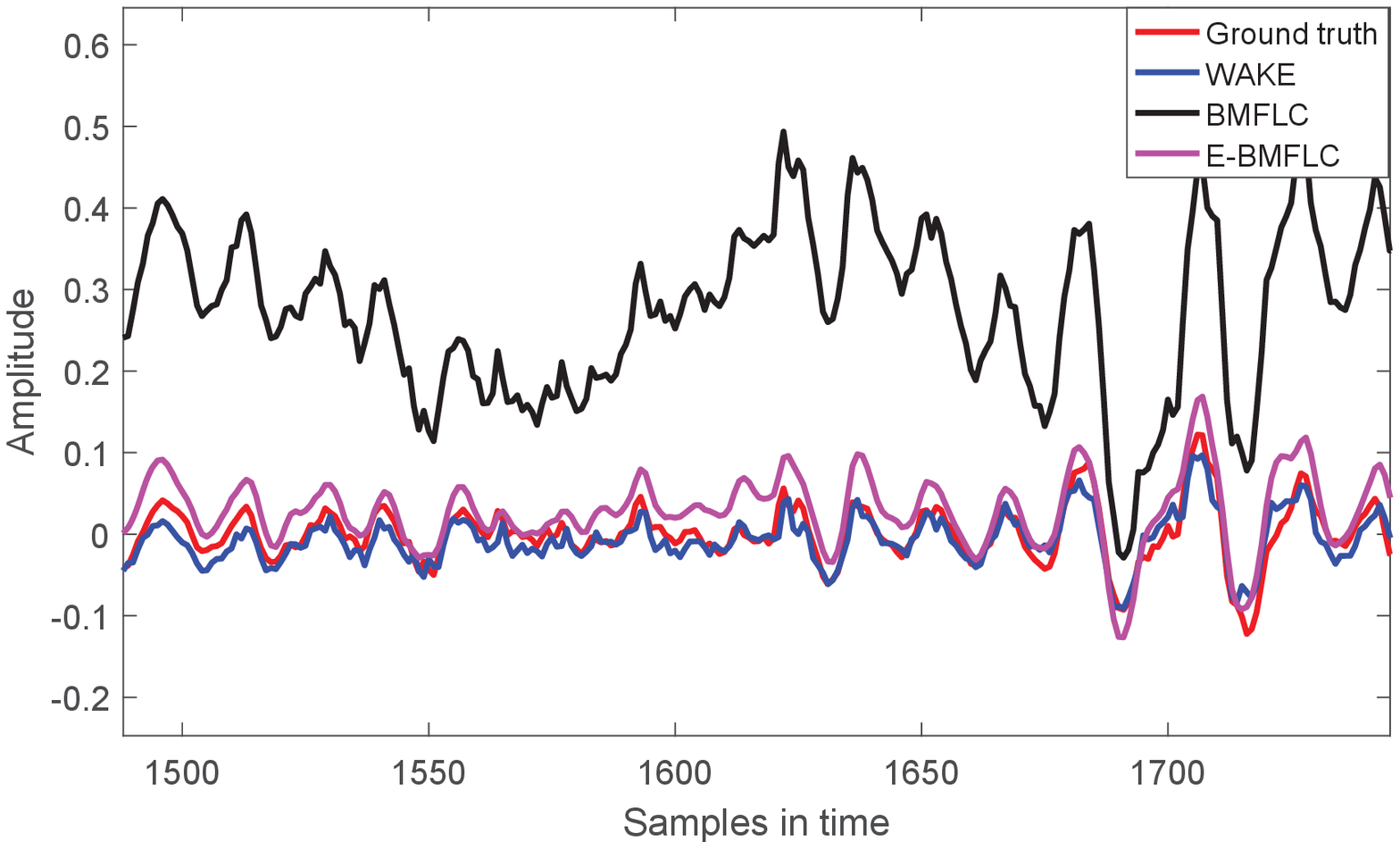}
	\caption{The predicted tremor derived by the proposed $\WAKE$ framework, the BMFLC method and the E-BMFLC method compared to the ground truth of the tremor derived by off-line processing of the signal (Subject 9 - $Y$-axis in Smartphone Dataset). \label{spain9}}
\end{figure}

\begin{table}
\centering
\footnotesize
\caption{Results of the BMFLC, E-BMFLC, and the proposed $\WAKE$ frameworks on the smartphone dataset. Term ``\textit{wname}'' in the table denotes the mother Wavelet function. 	\label{resultsSpain}}

\begin{tabular}{|c|c|c|c|c|c|c|c|c|c|}
\hline
\multirow{5}{*}{Subject} & \multicolumn{3}{|c|}{x-axis} & \multicolumn{3}{|c|}{y-axis} & \multicolumn{3}{|c|}{z-axis}\\
\cline{2-10}
&  \multicolumn{3}{|c|}{\textit{p=19, wname='db5'}} &  \multicolumn{3}{|c|}{\textit{p=15, wname='db7'}} &  \multicolumn{3}{|c|}{\textit{p=20, wname='db7'}} \\
&  \multicolumn{3}{|c|}{\textit{J=10, L=$2^{10}$}} &  \multicolumn{3}{|c|}{\textit{J=10, L=$2^{10}$}} &  \multicolumn{3}{|c|}{\textit{J=10, L=$2^{10}$}} \\
\cline{2-10}
& BMFLC & E-BMFLC& \WAKE & BMFLC & E-BMFLC& \WAKE & BMFLC & E-BMFLC& \WAKE  \\
\hline
1&.1284&.0895&\textbf{.0657}&.1506&.1038&\textbf{.0873}&.096&.1&\textbf{.064} \\
\hline
2&.0919&\textbf{.0728}&.0862&.1102&\textbf{.0640}&.0811&.0713&.0519&\textbf{.0433} \\
\hline
3&.0596&\textbf{.0366}&.0610&.1621&.0958&\textbf{.0511}&.0419&\textbf{.0291}&.0325 \\
\hline
4&.1093&.0767&\textbf{.0453}&.1112&.0746&\textbf{.0475}&.0754&.0599&\textbf{.0477} \\
\hline
5&.085&\textbf{.0648}&.078&.1015&.0574&\textbf{.0475}&.0844&.0528&\textbf{.0507} \\
\hline
6&.0637&.0638&\textbf{.0592}&.0647&.0391&\textbf{.0351}&.0661&.0427&\textbf{.0371} \\
\hline
7&.0881&.0499&\textbf{.0462}&.1892&.0877&\textbf{.0835}&.0815&.0501&\textbf{.0367} \\
\hline
8&.0629&\textbf{.0398}&.082&.1608&.0717&\textbf{.0349}&\textbf{.053}&.0716&.0914 \\
\hline
9&\textbf{.0419}&.0477&.0431&.1081&.0665&\textbf{.0321}&.0648&.068&\textbf{.052} \\
\hline
10&.0801&.0626&\textbf{.0563}&.0688&\textbf{.0455}&.0564&.0775&.0438&\textbf{.0431} \\
\hline		
\end{tabular}
\end{table}
}
%------------------------------------------------------------
%OOOOOOOOOOOOOOOOOOOOOOOOOOOOOOOOOOOOOOOOOOOOOOOOOOOOOOOOO
\section{Conclusion and Future Directions}  \label{sec:conc}
%OOOOOOOOOOOOOOOOOOOOOOOOOOOOOOOOOOOOOOOOOOOOOOOOOOOOOOOOO
In this paper, we proposed an adaptive auto-adjustable tremor extraction/estimation framework, referred to as $\WAKE$, which is an on-line, optimized, and multi-rate adaptive framework. The $\WAKE$ approach is developed based on wavelet decomposition and Kalman filtering to first estimate the voluntary motion in a myopic fashion (one-step ahead) and then use this information to extract involuntary movement (tremor). The low-frequency nature of voluntary component allows for an accurate myopic prediction which significantly enhances the performance of the technique. The $\WAKE$ framework, extracts a number of optimized hyper-parameters based on the behavior of the tremor which are then incorporated to boost the accuracy of estimation. The proposed $\WAKE$ framework is evaluated based on a designed synthetic dataset, and two real datasets namely called the Motus dataset and the Smartphone dataset. It is shown that the proposed approach outperforms its counterparts and provides superior performance even when compared to off-line techniques.

%OOOOOOOOOOOOOOOOOOOOOOOOOOOOOOOOOOOOOOOOOOOOOOOOOOOOOOOOO
%\newpage%\vspace{.2in}
\noindent
\textbf{Acknowledgment} \label{sec:ack}
%OOOOOOOOOOOOOOOOOOOOOOOOOOOOOOOOOOOOOOOOOOOOOOOOOOOOOOOOO

\noindent The authors would like to thank the ``\textit{Hospital Clinic de Barcelona}'' for providing the smartphone dataset.
It is worth mentioning that this work was partially supported by the Fonds de Recherche du Qu\'ebec Nature et Technologies (FRQNT) Grant 2018-NC-206591.

\bibliographystyle{elsarticle-num}
%OOOOOOOOOOOOOOOOOOOOOOOOOOOOOOOOOOOOOOOOOOOOOOOOOOOOOOOOO
\vspace{.1in}
\noindent
\textbf{References}
%OOOOOOOOOOOOOOOOOOOOOOOOOOOOOOOOOOOOOOOOOOOOOOOOOOOOOOOOO

\end{document}